\documentclass[aps,prd,amsmath,amssymb,preprintnumbers,nofootinbib,preprint,superscriptaddress]{revtex4-1}
\pdfoutput=1
\usepackage[utf8]{inputenc} 
\usepackage{graphicx}
\graphicspath{{./figures/}}
\usepackage{url}
\usepackage[bookmarks, pagebackref=false]{hyperref}
\usepackage[usenames,dvipsnames]{xcolor}
\definecolor{orange}{cmyk}{0,0.5,1,0}
\definecolor{rossoCP3}{cmyk}{0,.88,.77,.40}
\definecolor{graa}{rgb}{0.8,0.8,0.8}
\definecolor{blaa}{rgb}{0.2,0.2,0.6}
\hypersetup{
	colorlinks, 
	bookmarksopen, 
	bookmarksnumbered,
	citecolor=blaa, 		
	linkcolor=rossoCP3,	
	urlcolor=rossoCP3,			
	    }

\usepackage{amsthm}
\usepackage{bm}
\usepackage{bbm}
\usepackage{pxfonts}

\usepackage{amsmath,amssymb,amsfonts}
\usepackage{color}
\usepackage{float}
\usepackage{hyperref}
\usepackage[Symbolsmallscale]{upgreek}
\usepackage{amsmath}
\usepackage{amsfonts}
\usepackage{amssymb,dsfont}
\usepackage{graphicx}
\usepackage{amssymb}
\usepackage[vcentermath]{youngtab}
\usepackage[all]{xy}
\usepackage{pstricks}
\usepackage{dsfont}%
\usepackage{cases}
\usepackage{comment}

\usepackage{placeins}
\usepackage{xspace}
\usepackage{cancel} 

\usepackage{slashed}
\usepackage[caption=false, labelformat=simple, listofformat=subsimple, labelfont=default, margin=5pt,
    justification=raggedright]{subfig}

\makeatletter
	\renewcommand{\p@subfigure}{}
\makeatother

\usepackage{natbib}

\usepackage{feynmf}

\newcommand{\beq}{\begin{eqnarray}}
\newcommand{\eeq}{\end{eqnarray}}

\newcommand{\bmp}{\noindent\begin{minipage}{16cm}}
\newcommand{\emp}{\end{minipage}\vskip 7mm} 

\def\lsim{\mathrel{\rlap{\lower4pt\hbox{\hskip1pt$\sim$}}
    \raise1pt\hbox{$<$}}}                
\def\gsim{\mathrel{\rlap{\lower4pt\hbox{\hskip1pt$\sim$}}
    \raise1pt\hbox{$>$}}}                

\begin{document}

\title{\texorpdfstring{\large\color{rossoCP3}   Conformal Phase Diagram of Complete Asymptotically Free Theories}{Complete Asymptotic Freedom}}
\author{Claudio {\sc Pica}}
\email{pica@cp3.sdu.dk} 
\author{Thomas {\sc A. Ryttov}}
\email{ryttov@cp3.sdu.dk}
\author{Francesco {\sc Sannino}}
\email{sannino@cp3.sdu.dk}
\affiliation{{\color{rossoCP3} {CP}$^{ \bf 3}${-Origins}} \& the Danish Institute for Advanced Study {\color{rossoCP3}\rm{Danish IAS}},  University of Southern Denmark, Campusvej 55, DK-5230 Odense M, Denmark.}
\begin{abstract}
We  investigate the ultraviolet and infrared fixed point structure of gauge-Yukawa theories featuring a single gauge coupling, Yukawa coupling and scalar self coupling. Our investigations are performed using the two loop gauge beta function, one loop Yukawa beta function and one loop scalar beta function. We provide the general conditions that the beta function coefficients must abide for the theory to be completely asymptotically free while simultaneously possessing an infrared stable fixed point. We also uncover special trajectories in coupling space along which some couplings are both asymptotically safe and infrared conformal.
 \\
[.1cm]
{\footnotesize  \it Preprint: CP3-Origins-2016-025 DNRF90 }
\end{abstract}
\maketitle
\newpage
 
\section{Introduction}  
 
The discovery of the Higgs-like particle, so far, crowns the Standard Model (SM), a gauge-Yukawa theory, as one of the most successful theories of nature.  It is therefore imperative to gain vital  information about these fascinating theories. 
  
A natural classification of gauge-Yukawa theories can be made according to whether they admit  UV complete (in all the couplings) fixed points or they abide the full set of compositeness conditions. The presence of a UV fixed point guarantees the fundamentality of the theory since, setting aside gravity, it means that the theory is valid at arbitrary short distances.  If, however, a given gauge-Yukawa theory fails to be fundamental it can describe a composite theory in disguise provided it abides a set of UV compositeness conditions  \cite{Krog:2015bca}. In this limit it is a gauge theory augmented by four-fermion interactions \cite{Krog:2015bca}. 

If the UV fixed point occurs for vanishing values of the couplings the interactions are asymptotically free in the UV  \cite{Gross:1973id,Politzer:1973fx}. The fixed point is approached logarithmically and therefore, at short distances, perturbation theory is applicable. Asymptotic freedom is a UV phenomenon that still allows for several intriguing possibilities in the IR, depending on the specific underlying theory \cite{Sannino:2009za}. At low energies, for example, another interacting fixed point can occur. In this case the theory displays both long and short distance conformality. However the theory is interacting at large distances and the IR spectrum of the theory is continuous \cite{Georgi:2007ek}. Another possibility that can occur in the IR is that a dynamical mass is generated leading to either confinement or chiral symmetry breaking, or both. Certain subsets of  theories including nonsupersymmetric vector-like  fermionic gauge theories \cite{Caswell:1974gg,Banks:1981nn,Dietrich:2006cm,Sannino:2004qp,Pica:2010xq,Ryttov:2010iz,Mojaza:2012zd,Shrock:2015owa,Shrock:2013pya,Shrock:2013ca,Bergner:2015dya,Esbensen:2015cjw,Ryttov:2016hdp,Pica:2016rmv} \footnote{The first analysis of the conformal properties of baryonic operators for nonsupersymmetric gauge theories relevant for  popular extensions of the standard model has been performed in \cite{Pica:2016rmv} while in \cite{Ryttov:2016hdp} a consistent perturbative and scheme independent method for calculating such quantities has been studied. }, chiral gauge theories \cite{Appelquist:2000qg,Shi:2015fna,Shi:2015baa,Shi:2014yxa}, gauge-Yukawa theories \cite{Mojaza:2011rw,Grinstein:2011dq,Antipin:2011aa,Molgaard:2014mqa,Krog:2015bca} and purely scalar theories \cite{Shrock:2014zca} have been investigated in the literature.   Long overdue is, however, a more general and systematic classification of the dynamics of the gauge-Yukawa theories that begun in \cite{Antipin:2013sga}. 

An interesting  class of gauge-Yukawa theories are those displaying asymptotic freedom in all couplings. These are known as {\it complete asymptotically free} theories \cite{Gross:1973ju,Cheng:1973nv,Callaway:1988ya}.  Here the ultraviolet dynamics of Yukawa and scalar interactions is tamed by asymptotically free gauge fields; see \cite{Holdom:2014hla,Giudice:2014tma} for recent studies.  This phenomenon is quite distinct  from the recently discovered setup of {\it complete asymptotic safety }  \cite{Litim:2014uca,Litim:2015iea}. Here  the theory was found to flow to a nontrivial ultraviolet stable fixed point in a completely controllable manner \cite{Litim:2014uca,Litim:2015iea}.  The result shows that no additional symmetry principles, such as space-time supersymmetry \cite{Bagger:1990qh}, are required to ensure well-defined and predictive ultraviolet theories. Intriguingly {\it complete asymptotic safety} it is not a feature, in perturbation theory, of the gauge theory with either pure fermionic or scalar matter. Neither does the ultraviolet fixed point exist for the supersymmetrized version \cite{Intriligator:2015xxa,Martin:2000cr}.  Tantalising indications that ultraviolet interacting fixed point may exist nonperturbatively, and without the need of elementary scalars, appeared in \cite{Pica:2010xq}, and they were further explored in  \cite{Shrock:2013cca,Litim:2014uca} \footnote{Nonperturbative techniques are needed to establish the existence of such a fixed point when the number of colors and flavours is taken to be three and the number of UV light flavours is large but finite.  Asymptotic safety was originally introduced by Weinberg \cite{Weinberg:1980gg}  to address quantum aspects of  gravity  \cite{Litim:2011cp,Litim:2006dx,Niedermaier:2006ns,Niedermaier:2006wt,
Percacci:2007sz,Litim:2008tt,Reuter:2012id,Dona:2015tnf}. }.  Exciting possibilities for asymptotically safe extensions of the standard model include: the possibility that QCD itself could be completed at higher energies by a safe extension \cite{Sannino:2015sel};   asymptotically dafe dark matter model building \cite{Sannino:2014lxa} and asymptotically safe inflation \cite{Nielsen:2015una,Svendsen:2016kvn}.

These observations should make it clear that further studies of gauge-Yukawa theories are to be carried out. Here we take one step further and explore, within perturbation theory, the novel infrared structure of  a wide class of {\it complete asymptotically free} theories \cite{Gross:1973ju,Cheng:1973nv,Callaway:1988ya}.  

The paper is organised as follows. In Section \ref{one} we review the conditions for complete asymptotic freedom for a wide class of gauge theories also investigated in \cite{Callaway:1988ya,Holdom:2014hla,Giudice:2014tma}. The entirely novel part of our work resides in Sections \ref{tre} and \ref{quattro}. Here we first derive the general conditions for the presence of interacting fixed points in all couplings and then classify them. We conclude with the general conformal phase diagram of complete asymptotically safe quantum field theories.

\section{Complete Asymptotic Freedom Conditions}
\label{one}

Since we are interested in the perturbatively calculable structure of the phase diagram we will consider generic Gauge-Yukawa theories with scalars that are also gauged under the gauge group. We will focus on three marginal couplings at the classical level, i.e. the gauge, the Yukawa and a scalar self coupling defined as follows: 
\begin{equation}
	\alpha_g = \frac{g^2}{(4 \pi)^2},
	\quad
	\alpha_y = \frac{y^2}{(4 \pi)^2},
	\quad
	\alpha_\lambda = \frac{\lambda}{(4 \pi)^2}.
	\label{eq:alphas}
\end{equation}
We first start with reviewing the conditions for the presence of complete asymptotic freedom and then go beyond the state-of-the-art by systematically investigating  the possible infrared conformal structure of these theories. 
\subsection{Gauge and Yukawa subsystem}

We begin by first investigating the pure gauge system with a single gauge coupling and no Yukawa and self couplings. To one loop order the running of the gauge coupling is dictated by the following renormalization group equation
\begin{eqnarray}\label{gauge}
\mu \frac{d \alpha_g}{d \mu} &=& b_0 \alpha_g^2
\end{eqnarray}
Note that it has a degenerate fixed point at the origin. Its solution is simply
\begin{eqnarray}
\alpha_g &=& \frac{\alpha_{g0}}{1-b_0 \alpha_{g0}\ln \frac{\mu}{\mu_0}}  
\end{eqnarray}
where $\alpha_{g0}=\alpha_g(\mu_0)$ and $\mu_0$ is some fixed scale. If we choose
\begin{eqnarray}\label{AF1}
b_0<0
\end{eqnarray}
then the theory is asymptotically free. In the deep ultraviolet the coupling approaches the trivial fixed point and vanishes. Also note that technically the solution to the running coupling also contains an unphysical branch below the scale $\mu_0 \exp \left[ \frac{1}{b_0 \alpha_{g0}} \right]$. Here the coupling is negative and approaches the trivial fixed point in the deep infrared. 

We now continue this section by adding to the pure gauge system also a Yukawa coupling. To one loop order the renormalization group equation for the Yukawa coupling is 
\begin{eqnarray}\label{Yukawa}
\mu \frac{d \alpha_H}{d \mu} &=&  \alpha_H \left[ c_1 \alpha_g + c_2 \alpha_H \right] 
\end{eqnarray}
where in general $c_1<0$ and $c_2>0$. Consider first the simplest case in which there is no gauge coupling. Here the running of the Yukawa is easily found to be 
\begin{eqnarray}
\alpha_H &=& \frac{\alpha_{H0}}{1-c_1 \alpha_{H0}\ln \frac{\mu}{\mu_0}}
\end{eqnarray} 
Again technically there are two branches to the running of the coupling. In the deep infrared the coupling flows to the trivial fixed point while at larger scales there is a Landau pole. This is the physical branch since here the coupling is positive. Beyond the Landau pole the coupling is negative while approaching the trivial fixed point in the deep ultraviolet. This is an unphysical branch. Hence a single Yukawa coupling on its own and without the contribution of any other couplings can never be asymptotically free.

Switching on the gauge coupling we now must solve the coupled set of renormalization group equations, Eq. \ref{gauge} and \ref{Yukawa}. In order to do this we will first combine the two equations by forming the ratio $\frac{\beta_H}{\beta_g}$ to obtain
\begin{eqnarray}
\frac{d \alpha_H}{d \alpha_g} &=& \frac{1}{b_0} \frac{\alpha_H}{\alpha_g} \left( c_1 + c_2 \frac{\alpha_H}{\alpha_g} \right)
\end{eqnarray}
The solution to this equation is
\begin{eqnarray}
\alpha_H &=&\frac{\alpha_{H0}}{ \left(1 - \frac{c_2}{b_0-c_1} \frac{\alpha_{H0}}{\alpha_{g0}} \right) \alpha_{g0}^{\frac{c_1}{b_0}}  \alpha_{g}^{-\frac{c_1}{b_0}+1} + \frac{c_2}{b_0-c_1} \alpha_{H0} } \alpha_g \ , \qquad b_0 \neq c_1 \\
\alpha_{H} &=& \frac{\alpha_{H0}}{\left(1 + \frac{c_2}{c_1} \frac{\alpha_{H0}}{\alpha_{g0}} \ln \alpha_{g0} \right) \alpha_{g0}  - \frac{c_2}{c_1} \alpha_{H0} \ln \alpha_g} \alpha_g \ , \qquad b_0 = c_1
\end{eqnarray}
Hence the specific solution for the running of the Yukawa coupling depends on the values of $b_0$ and $c_1$. We are searching for solutions where the Yukawa coupling is asymptotically free. This implies that the Yukawa coupling must be positive, vanish asymptotically and contain no Landau poles. Landau poles could potentially show up if the denominator vanishes for some value of the gauge coupling. 

First we quickly discard the situation with $b_0=c_1$. As the gauge coupling decreases asymptotically, barring any potential Landau poles, the logarithmic term in the denominator dominates and the Yukawa coupling tends to zero. However since the coefficient $\frac{c_2}{c_1}<0$ is always negative so is the Yukawa coupling asymptotically. Hence it cannot be asymptotically free.

We will now examine the more interesting case with $b_0\neq c_1$. There are two terms in the denominator contributing to the running of the Yukawa coupling as the gauge coupling decrease. If $-\frac{c_1}{b_0}+1 <0$ the first term dominates while if $-\frac{c_1}{b_0}+1 >0$ the second term dominates. The former constraint is equivalent to $b_0-c_1>0$ while the latter constraint corresponds to $b_0-c_1<0$. Note that the last option $-\frac{c_1}{b_0}+1 =0$ has been excluded above. 

In the latter case the Yukawa coupling scales as $\frac{c_2}{b_0-c_1}\alpha_g$ and hence is negative. We therefore exclude this possibility. In the former case however the Yukawa coupling can be asymptotically free provided that the coefficient in front of the leading term in the denominator is positive. Hence in this case
\begin{eqnarray}\label{gY}
\alpha_H &\sim& \frac{\alpha_{H0}}{\left(1-\frac{c_2}{b_0-c_1} \frac{\alpha_{H0}}{\alpha_{g0}} \right) \alpha_{g0}^{\frac{c_1}{b_0}}} \alpha_{g}^{\frac{c_1}{b_0}} \ , \qquad \frac{\alpha_{g0}}{\alpha_{H0}} > \frac{c_2}{b_0-c_1} \ , \qquad b_0-c_1>0
\end{eqnarray}
Note that the Yukawa coupling tends to zero faster that the gauge coupling. Also whether or not the Yukawa coupling is asymptotically free depends on the values of couplings in the infrared. There is also the special case for which the coefficient of the first term in the denominator vanishes, i.e. where the values of the couplings in the infrared are fine tuned. Here the Yukawa coupling scales as the gauge coupling with
\begin{eqnarray}\label{gYfixed}
\alpha_{H} &=& \frac{b_0-c_1}{c_2} \alpha_g \ , \qquad \frac{\alpha_{g0}}{\alpha_{H0}} = \frac{c_2}{b_0-c_1}  \ , \qquad b_0-c_1>0
\end{eqnarray} 
Following \cite{Giudice:2014tma} we will refer to this case as fixed flow. Lastly there is the possibility where the values of the couplings in the infrared do not satisfy the above constraints. This corresponds exactly to the case where the Yukawa coupling develops a Landau pole. Namely here the value of the gauge coupling at the zero of the denominator is positive and hence the Yukawa is bound to diverge as the gauge coupling decreases and reaches this critical value.

\begin{figure}
\includegraphics[width=0.5\textwidth]{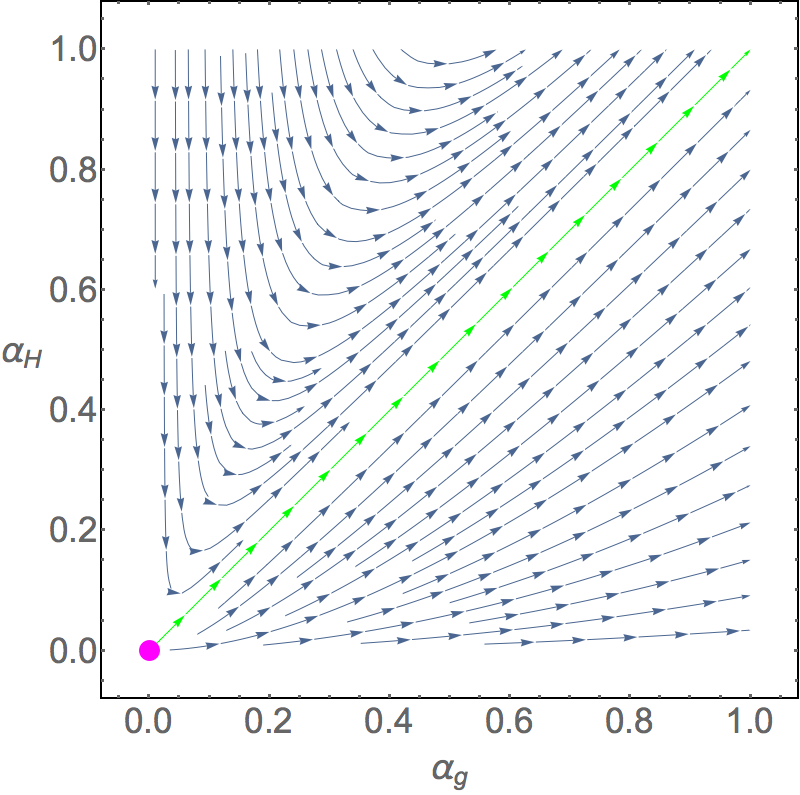}
\caption{Flow of the gauge and Yukawa couplings near the trivial UV fixed point for $b_0=-1$, $c_1 = -2$ and $c_2=1$. The green flow trajectory is the fixed flow where the gauge and Yukawa couplings scale the same way, Eq. \ref{gYfixed}. Below the fixed flow trajectory are the trajectories for the remaining asymptotically free theories, Eq. \ref{gY}} \label{fig:gY1loop}
\end{figure}

In Fig. \ref{fig:gY1loop} we plot the flow of the couplings for a set of representative values of $b_0,c_1,c_2$. The green trajectory corresponds to the fixed flow where the couplings scale proportionally. Below the green trajectory the theory is asymptotically free but with the Yukawa coupling vanishing faster than the gauge coupling. Lastly above the green trajectory the Yukawa coupling develops a Landau pole. Which trajectory the system follows depend on the fixed values of the couplings in the infrared.

\subsection{The scalar self-interactions}

To one loop order the beta function of a single self coupling is
\begin{eqnarray}
\mu \frac{d\alpha_{\lambda}}{d\mu} &=& \alpha_{\lambda} \left( d_1\alpha_{\lambda} + d_2 \alpha_g + d_3 \alpha_H  \right) + d_4 \alpha_g^2 + d_5 \alpha_H^2
\end{eqnarray}
where $d_1,d_3,d_4\geq 0$ and $d_2,d_5\leq 0$. Together with Eq. \ref{gauge} and \ref{Yukawa} it describes the running of the gauge, Yukawa and self coupling in a general gauge-Yukawa system at one loop order. We begin slowly by first investigating the behaviour of the self coupling assuming the absence of both the gauge and Yukawa couplings (corresponding to $d_2=d_3=d_4=d_5=0$). Here the solution to the renormalisation group equation is 
\begin{eqnarray}
\alpha_{\lambda} &=& \frac{\alpha_{\lambda0}}{1-d_1\alpha_{\lambda 0}\ln \frac{\mu}{\mu_0}}
\end{eqnarray}
First note that in this approximation the beta function has a degenerate fixed point at the origin. Since $d_1 \geq 0$ the self coupling flows to the trivial fixed point in the deep infrared while it exhibits a Landau pole at the scale $\mu_0 \exp \left[\frac{1}{d_1 \alpha_{\lambda0}} \right]$. There is also a second branch which lies beyond the scale of the Landau pole. Here the self coupling flows to the trivial fixed point in the deep ultraviolet. However the value of the coupling is negative and hence this situation must be discarded since the system is unstable. 

We will now turn our attention to the general case with gauge and Yukawa couplings included. We shall not attempt to solve the renormalization group equations in generality but only solve it asymptotically in the large energy region. Forming first the ratio $\frac{\beta_{\lambda}}{\beta_g}$ we obtain
\begin{eqnarray}
\frac{d \alpha_{\lambda}}{d \alpha_g} &=& \frac{1}{b_0} \frac{\alpha_{\lambda}}{\alpha_g} \left( d_1 \frac{\alpha_{\lambda}}{\alpha_g} + d_2 + d_3 \frac{\alpha_H}{\alpha_g}  \right) + \frac{d_4}{b_0} +\frac{d_5}{b_0} \left(  \frac{\alpha_H}{\alpha_g} \right)^2
\end{eqnarray}
We shall distinguish between whether the gauge and Yukawa couplings are on a fixed flow or not. First if we assume that they are not on their fixed flow then asymptotically we found above that the Yukawa coupling tends to zero faster than the gauge copuling. Hence we can simply look for a solution to the differential equation
\begin{eqnarray}
\frac{d \alpha_{\lambda}}{d \alpha_g} &=& \frac{1}{b_0} \frac{\alpha_{\lambda}}{\alpha_g} \left( d_1 \frac{\alpha_{\lambda}}{\alpha_g} + d_2  \right) + \frac{d_4}{b_0} 
\end{eqnarray}
Note that the Yukawa coupling has disappeared from the equation. If the gauge and Yukawa couplings are not on their fixed flow the asymptotic running of the self coupling does not depend on the Yukawa coupling. The solution to the above differential equation can be written as
\begin{eqnarray}
\alpha_{\lambda} &=&  \left( b_0 - d_2 + \sqrt{-k} \tan \left[ - \text{arctan} \left( \frac{k_0  }{\sqrt{-k} \alpha_{g0} }\right) + \frac{\sqrt{-k}}{2b_0} \ln \frac{\alpha_g}{\alpha_{g0}}   \right]  \right) \frac{\alpha_g}{2d_1}\ ,  \qquad k<0  \\
\alpha_{\lambda} &=& \left( b_0 - d_2 -\sqrt{k} \frac{k_0+\sqrt{k}\alpha_{g0} +\left( k_0-\sqrt{k}\alpha_{g0} \right) \frac{\alpha_g}{\alpha_{g0}}^{-\frac{\sqrt{k}}{b_0}} }{k_0+\sqrt{k}\alpha_{g0} - \left( k_0-\sqrt{k}\alpha_{g0} \right) \frac{\alpha_g}{\alpha_{g0}}^{-\frac{\sqrt{k}}{b_0}}}  \right) \frac{\alpha_g}{2d_1} \ , \qquad k>0 \\
\alpha_{\lambda} &=& \frac{4b_0d_1 \alpha_{\lambda 0} + k_0 (b_0-d_2)  \ln \frac{\alpha_g}{\alpha_{g0}} }{2b_0 \alpha_{g0} +k_0 \ln \frac{\alpha_g}{\alpha_{g0}} } \frac{\alpha_g}{2d_1}   \ , \qquad k=0
\end{eqnarray} 
with 
\begin{eqnarray}\label{eq:k}
k &=&  (b_0 - d_2)^2 - 4 d_1d_4 \\
k_0 &=& \left(b_0 - d_2  \right)\alpha_{g0} - 2d_1 \alpha_{\lambda 0}
\end{eqnarray}
We need to understand the behaviour of this solution asymptotically as the gauge coupling decreases from its value in the infrared. Consider first the case $k<0$. As we continuously vary the gauge coupling the self coupling will diverge due to the periodicity in tangent. Hence it will inevitably lead to the existence of Landau poles in the self coupling and we therefore discard this possibility.

Instead take $k>0$. First consider the two limiting cases with $k_0 + \sqrt{k} \alpha_{g0}=0 $ or $k_0 - \sqrt{k}\alpha_{g0}=0$. Here the self and gauge couplings are on a fixed flow with 
\begin{eqnarray}\label{eq:gY}
\alpha_{\lambda} &=& \frac{b_0 - d_2 + \sqrt{k}}{2d_1}\alpha_g \qquad \text{or} \qquad \alpha_{\lambda} =   \frac{b_0 - d_2 - \sqrt{k}}{2d_1}\alpha_g 
\end{eqnarray}
Positivity of the self coupling then implies that it is asymptotically free along both directions if $ \frac{b_0 - d_2 - \sqrt{k}}{2d_1} >0$, asymptotically free along only the first direction if $ \frac{b_0 - d_2 + \sqrt{k}}{2d_1} >0$ and $ \frac{b_0 - d_2 - \sqrt{k}}{2d_1} <0$ and non-asymptotically free along both directions if $ \frac{b_0 - d_2 + \sqrt{k}}{2d_1} <0$. These two limiting cases trace two straight trajectories in the gauge and self coupling plane. 

Since we want to know whether there are other trajectories along which the self coupling is asymptotically free we must make sure that it has no poles as we vary the gauge coupling. Poles will show up if the denominator vanishes for a non-negative value of the gauge coupling, i.e. if the following equation has a solution 
\begin{eqnarray}\label{zero}
\left( \frac{\alpha_g}{\alpha_{g0}} \right)^{- \frac{\sqrt{k}}{b_0} } &=&  \frac{k_0 +\sqrt{k} \alpha_{g0}}{k_0 - \sqrt{k} \alpha_{g0}}
\end{eqnarray}
for some $\alpha_g \geq 0$. We can quickly discard the case with a vanishing value of the gauge coupling since this would require $k_0 + \sqrt{k} \alpha_{g0}=0$ which we saw above leads to the self coupling and gauge coupling being on a fixed flow. In other words the zero is not there being cancelled by a zero in the numerator. Poles however could exist for positive values of the gauge coupling. This would require that the right hand side of Eq. \ref{zero} be larger than zero implying that both its numerator and denominator must be of the same sign. Therefore since $- \frac{\sqrt{k}}{b_0} >0 $ there exists a pole in the self coupling in the ultraviolet (corresponding to $\alpha_g<\alpha_{g0}$) if the right hand side is less than unity and a pole in the self coupling in the infrared (corresponding to $\alpha_g>\alpha_{g0}$) if the right hand side is larger than unity.

First we will worry about an eventual Landau pole in the ultraviolet. As noted above for this to happen the right hand side should be less than unity. Since the denominator is always less than the numerator (but should be of the same sign) this can only occur provided the numerator is less than zero. Hence the self coupling will contain a Landau pole in the ultraviolet unless the numerator is positive $k_0 + \sqrt{k} \alpha_{g0}>0$ corresponding to the following condition 
\begin{eqnarray}\label{eq:1}
\frac{\alpha_{\lambda 0}}{\alpha_{g0}} &<& \frac{b_0-d_2 + \sqrt{k}}{2d_1}
\end{eqnarray}
If this condition is satisfied by the fixed values $\alpha_{g0}$ and $\alpha_{\lambda 0}$ the self coupling contains no Landau poles in the ultraviolet and it vanishes asymptotically together with the gauge coupling.

\begin{figure}[h]
\includegraphics[width=0.5\textwidth]{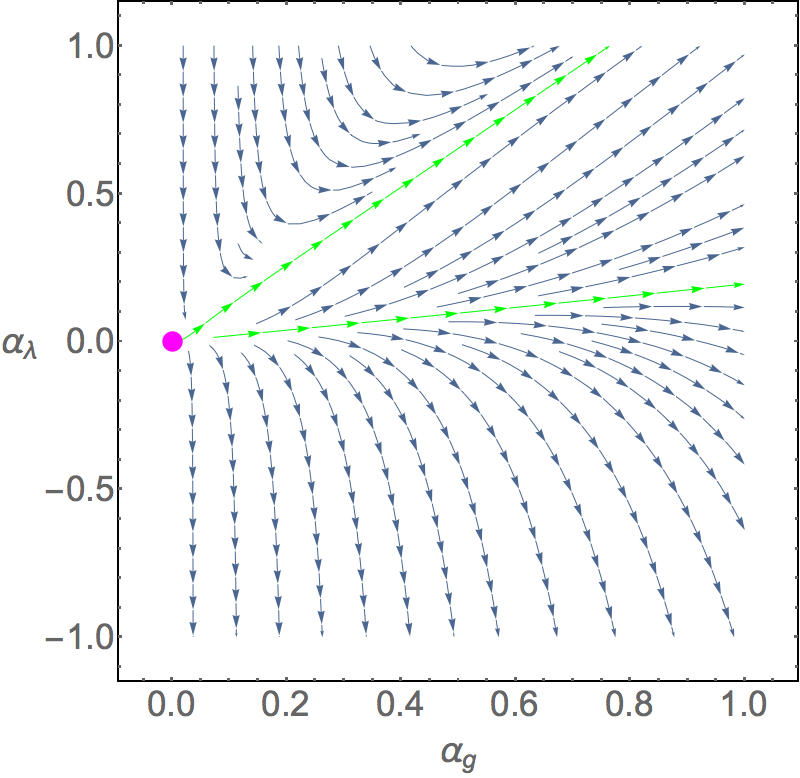}
\caption{Flow of the gauge and self couplings for $b_0=-1$, $d_1=2$, $d_2=-4$ and $d_4=\frac{1}{2}$. This choice renders $k=5$. The flow is illustrated at one loop in both the gauge and self couplings. The violet point is the ultraviolet fixed point. } \label{fig:gl1loop}
\end{figure}

As noted above there might also be a pole in the self coupling in the infrared. At first it might seem that we shouldn't really worry about them since we are only interested in asymptotically free theories. However these infrared poles occur for a negative value of the self coupling destabilising the system. This must necessarily be so since for an infrared pole to exist the right hand side of Eq. \ref{zero} must be larger than unity implying that denominator $k_0 - \sqrt{k}\alpha_{g0}>0$ must be positive. Hence as we increase the gauge coupling from its fixed value $\alpha_{g0}$ towards the pole the self coupling grows to negative infinity. Therefore for the self coupling not to have any negative poles in the infrared we must demand that $k_0 - \sqrt{k}\alpha_{g0}<0$ corresponding to
\begin{eqnarray}\label{eq:2}
\frac{\alpha_{\lambda 0}}{\alpha_{g0}} &>& \frac{b_0 - d_2 - \sqrt{k}}{2d_1}
\end{eqnarray}
Therefore within the window marked by the conditions Eq. \ref{eq:gY}, \ref{eq:1}, \ref{eq:2}, namely
\begin{eqnarray}
\frac{b_0 - d_2 - \sqrt{k}}{2d_1} & \leq & \frac{\alpha_{\lambda 0}}{\alpha_{g0}} \leq \frac{b_0 - d_2 + \sqrt{k}}{2d_1}
\end{eqnarray}
the self coupling flows to the trivial fixed point in the ultraviolet and has no negative poles in the infrared. On the boundaries of the window the gauge and self couplings are on a fixed flow. Lastly we need to make sure that at least a single trajectory within the window is for a positive value of the self coupling at all scales. This is not automatically satisfied but is ensured provided we take the upper end of the window to be positive. In other words at least a single trajectory along which the self coupling is asymptotically free exists if the following condittion
\begin{eqnarray}
b_0 - d_2 + \sqrt{k} &>&0
\end{eqnarray}
is satisfied. In Fig. \ref{fig:gl1loop} we plot the flow of the gauge and self couplings at one loop order under the assumption that $k>0$ and that the gauge and Yukawa couplings are not on their fixed flow. The green trajectories are the two exact fixed flow solutions Eq. \ref{eq:gY}. They mark the boundary within which the theory is asymptotically free.

\begin{figure}[h!]
\includegraphics[width=0.5\textwidth]{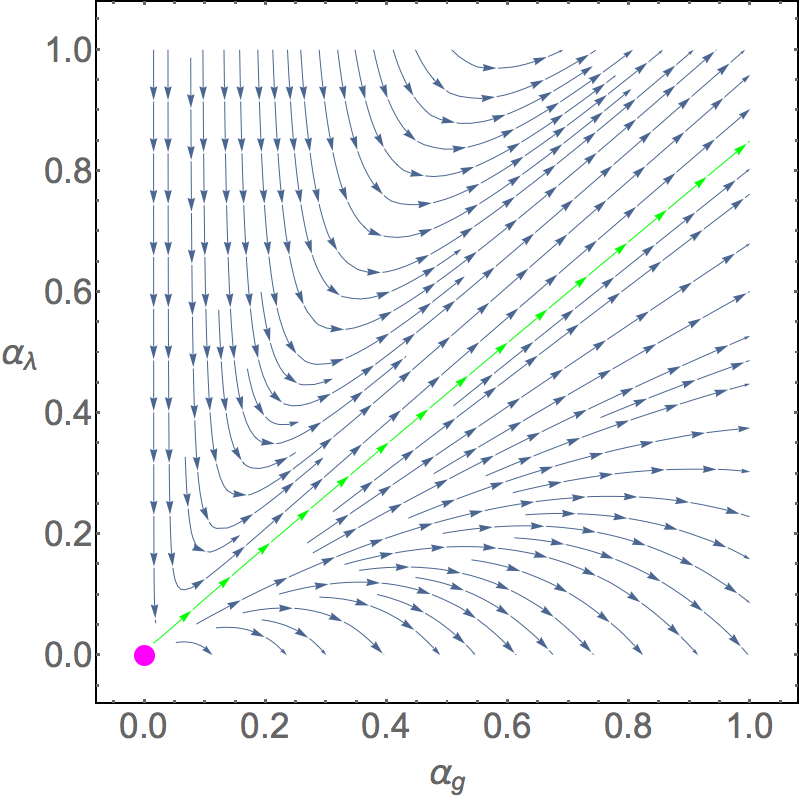}
\caption{Flow of the gauge and self couplings for $b_0=-1$, $d_1=2$, $d_2=-4$ and $d_4=\frac{9}{8}$. This choice renders $k=0$. The flow is illustrated at one loop in both the gauge and self couplings. The violet point is the ultraviolet fixed point. } \label{fig:gl1loop0}
\end{figure}

Lastly we need to discuss the special case $k=0$. Again we need to worry about potential poles appearing in the running of the self coupling. Poles will appear if the denominator vanishes for a non-negative value of the gauge coupling. In other words if there exists a solution to 
\begin{eqnarray}
\ln \frac{\alpha_g}{\alpha_{g0}} = - \frac{2b_0 \alpha_{g0}}{k_0}
\end{eqnarray}
for some $\alpha_g \geq 0$. If the fixed values $\alpha_{g0}$ and $\alpha_{\lambda 0}$ are chosen such that $k_0<0$ ($k_0>0$) then the self coupling has an ultraviolet (infrared) pole corresponding to $\alpha_g < \alpha_{g0}$ ($\alpha_g > \alpha_{g0}$). At the ultraviolet pole the self coupling plus to plus infinity while at the infrared pole the self coupling blows to minus infinity signalling an instability. It is therefore only possible for the self coupling to be asymptotically free if the fixed point values are chosen such that $k_0 = 0$. Here on this specific trajectory the gauge coupling and self coupling are on a fixed flow with
\begin{eqnarray}
\frac{\alpha_{\lambda}}{\alpha_{\lambda 0}} &=& \frac{\alpha_g}{\alpha_{g0}} \ , \qquad k_0=0
\end{eqnarray}
We plot the flow in Fig. \ref{fig:gl1loop0} for a representative set of values of beta function coefficients yielding $k=0$. It is only along the green trajectory that the self coupling is asymptotically free.

What we have learned so far is that asymptotic freedom of the self coupling is dictated by the value of $k$ which only depends on the values of certain beta function coefficients and the fixed values of the couplings at some scale. We can imagine scanning the parameter space of coefficients and hence varying $k$. If we increase the value of $k$ the window where the self coupling is asymptotically free opens up while it shrinks to a line for $k=0$. For negative $k$ the self coupling cannot be asymptotically free. 

What remains to be studied is the situation where the gauge and Yukawa couplings are on their fixed flow. Here $\alpha_H = \frac{b_0-c_1}{c_2}\alpha_g$ and we can write the differential equation that governs the running of the self coupling as
\begin{eqnarray}
\frac{d \alpha_{\lambda}}{d \alpha_g} &=& \frac{1}{b_0} \frac{\alpha_{\lambda}}{\alpha_g} \left( d_1 \frac{\alpha_{\lambda}}{\alpha_g}  + d_2 +d_3 \frac{b_0-c_1}{c_2} \right) + \frac{d_4}{b_0} + \frac{d_5}{b_0}\left( \frac{b_0-c_1}{c_2} \right)^2
\end{eqnarray}
One might have feared that not being able to neglect the running of the Yukawa coupling would have complicated the situation. However we see that the above differential equation has the same form as the differential equation governing the running of the self coupling in the absence of the Yukawa contribution. All we need to do is to replicate our analysis above and make the following substitutions 
\begin{eqnarray}\label{eq:d2'}
d_2 &\rightarrow&  d_2'= d_2 +d_3 \frac{b_0-c_1}{c_2} \\
d_4 &\rightarrow& d_4' = d_4 + d_5 \left( \frac{b_0-c_1}{c_2} \right)^2 \\
k &\rightarrow& k' =  \left( b_0 - d_2 -d_3 \frac{b_0-c_1}{c_2}  \right)^2 - 4d_1 \left(d_4 + d_5 \left( \frac{b_0-c_1}{c_2} \right)^2 \right) \label{eq:k'}
\end{eqnarray}
With these replacements the conclusions from above can be taken directly over to the case where the gauge and Yukawa couplings are on their fixed flow. Specifically within the window 
\begin{eqnarray}
\frac{b_0 - d_2' - \sqrt{k'}}{2d_1} \leq \frac{\alpha_{\lambda 0}}{\alpha_{g0}} \leq  \frac{b_0 - d_2' + \sqrt{k'}}{2d_1} \ , \qquad k'>0
\end{eqnarray}
the self coupling has no poles and as long as the upper boundary is positive $b_0 - d_2' + \sqrt{k'}>0$ there exists at least one trajectory along which it is asymptotically free. Finally we have the case with $k'=0$ for which there is a single trajectory along which the self coupling is asymptotically free with the self and gauge (and therefore also Yukawa) couplings being on a fixed flow with
\begin{eqnarray}
\frac{\alpha_g}{\alpha_{g0}}  &=& \frac{\alpha_H}{\alpha_{H0}} = \frac{\alpha_{\lambda}}{\alpha_{\lambda 0}} \ , \qquad k'=0
\end{eqnarray}

\subsection{Summary of the CAF conditions }

Here we briefly summarise the necessary conditions that the beta function coefficients must satisfy in order for all three couplings to be asymptotically free. If the gauge and Yukawa couplings are not on their fixed flow these conditions are
\begin{eqnarray}
b_0 <0  \ , \qquad b_0-c_1 >0 \ , \qquad k \geq 0 \ , \qquad b_0 - d_2 +\sqrt{k}>0 \ , \qquad \text{Condition CAF}_1
\end{eqnarray}
where $k$ is given by Eq. \ref{eq:k}. If the beta function coefficients satisfy these constraints and the couplings satisfy appropriate initial (infrared) conditions the theory is \emph{complete asymptotically free}. The first (second) condition is necessary to ensure asymptotic freedom of the gauge (Yukawa) coupling while the third and fourth conditions are necessary to ensure asymptotic freedom and positivity of the self coupling. 

On the other hand if the gauge and Yukawa couplings are on their fixed flow then the necessary set of conditions that the beta function coefficients must satisfy is
\begin{eqnarray}
b_0 <0  \ , \qquad b_0-c_1 >0 \ , \qquad k' \geq 0 \ , \qquad b_0 - d_2' + \sqrt{k'}>0 \ , \qquad \text{Condition CAF}_2
\end{eqnarray}
where $k'$ and $d_2'$ are given by Eq. \ref{eq:d2'} and \ref{eq:k'}. The condition for asymptotic freedom of the self coupling is in this case different from the condition where the gauge and Yukawa couplings are not on their fixed flow. This is because the running of the Yukawa coupling can no  longer be neglected and has an influence on the running of the self coupling. If these contions CAF$_2$ are satisfied and the couplings satisfy appropriate initial (infrared) conditions the theory is \emph{complete asymptotically free}.

\section{Interacting IR fixed point}
\label{tre}

In this section we shall study higher order corrections to the beta functions and thereby reveal a more complicated phase structure than the one visualised in Fig. \ref{fig:gY1loop}. In particular we will search for IR fixed points. In order to satisfy the Weyl consistency conditions we shall, as a first step, use two loops in the gauge coupling and one loop in the Yukawa coupling. To this order the beta functions read
\begin{eqnarray}
\beta_g &=&  \alpha_g^2 \left(b_0 + b_1 \alpha_g + b_H \alpha_H \right) \\
\beta_H &=& \alpha_H \left( c_1 \alpha_g + c_2 \alpha_H \right) 
\end{eqnarray}
Besides the trivial fixed point studied above the system now possesses the following (non)trivial fixed points
\begin{eqnarray}
\alpha_{g,1*} =  \frac{-b_0}{b_1} \ , \qquad \alpha_{H,1*} = 0 \ , \qquad \text{FP}_1
\end{eqnarray}
and
\begin{eqnarray}
\alpha_{g,2*} = \frac{-b_0}{b_1^{\text{eff}} } \ , \qquad \alpha_{H,2*} = \frac{{c_1}b_0 }{{c_2}b_1^{\text{eff}} } \ , \qquad \text{FP}_2
\end{eqnarray}
where $b_{1}^{\text{eff}} = b_1 - \frac{c_1}{c_2}b_H$. The first  fixed point FP$_1$ corresponds to the usual Banks-Zaks fixed point of the gauge coupling if the Yukawa coupling is switched off while the existence of the second nontrivial fixed point FP$_2$ is due to the interplay between both the gauge and Yukawa coupling. 

First we note that all nontrivial fixed point values are proportional to the first gauge beta function coefficient $b_0$. For a specific theory $b_0$ depends on the gauge fields as well as the matter content charged under the gauge symmetry. We can always imagine to pick a theory for which $b_0$ is arbitrarily close to zero making the above fixed points perturbative and therefore reliable provided the condition for asymptotic freedom $b_0-c_1>0$ is   satisfied.

For the fixed points to be physical we must require that they occur for a real and positive value of the $\alpha_g $ and $\alpha_H$ being they the squared of gauge and Yukawa couplings. This yields the conditions
\begin{eqnarray}
b_1>0  \  , \qquad \text{FP}_1 \\
  b_1^{\text{eff}} >0 \  , \qquad \text{FP}_2 
\end{eqnarray}
Note that since $b_H$ in principle can be both positive or negative the fixed points can exist simultaneously or independently of each other. 

In order to study the fixed points having included a self coupling we will first make use of the gauge beta function to two loops, the Yukawa and self coupling beta function to one loop. Even though this loop counting does not satisfy the Weyl consistency conditions it will make an easier start than jumping straight into a study of the three loop gauge beta function, two loop Yukawa beta function and one loop self coupling. The set of beta function we want to study is therefore
\begin{eqnarray}
\beta_g &=&  \alpha_g^2 \left(b_0 + b_1 \alpha_g + b_H \alpha_H \right) \\
\beta_H &=& \alpha_H \left( c_1 \alpha_g + c_2 \alpha_H \right)  \\
\beta_{\lambda} &=& \alpha_{\lambda} \left( d_1\alpha_{\lambda} + d_2 \alpha_g + d_3 \alpha_H  \right) + d_4 \alpha_g^2 + d_5 \alpha_H^2
\end{eqnarray}
Since the gauge and Yukawa beta functions do not depend on the self coupling to this order the fixed points are the same as in the case without a self coupling. We then need to set to zero the self coupling beta function in order to find the fixed point solutions for the self coupling. Doing so we find
\begin{eqnarray}
\alpha_{g,1*} =  \frac{-b_0}{b_1} \ , \qquad \alpha_{H,1*} = 0 \ , \qquad \alpha_{\lambda,1*}^{\pm} &=& \frac{b_0 \left( d_2\pm \sqrt{d_2^2 - 4d_1d_4} \right) }{2b_1d_1} \ , \qquad \text{FP}_1^{\pm}
\end{eqnarray}
where $\pm$ refers to the two possible zeros for the self coupling fixed point. Whether they are physical depends on the values of the beta function coefficients. There exists either none, one or two positive fixed points. Assuming that all three couplings are asymptotically free they are positive provided 
\begin{eqnarray}\label{eq:FP1}
b_1>0   \ , \qquad  l_1 = d_2^2 - 4d_1d_4 \geq 0
\end{eqnarray}
Note that if one fixed point is positive then the other is also bound to be positive. One fixed point cannot be positive while the other is negative and vice versa assuming asymptotic freedom and positivity of $\alpha_{g*}$. There is also the special case where $l_1=0$ and the fixed point solutions collapse to a single solution FP$_1^{+}=$FP$_1^{-}$. Note that if these two fixed points exist then they exist independently of whether the gauge and Yukawa couplings are on their fixed flow or not, i.e. independent of whether the conditions CAF$_1$ or CAF$_2$ are satisfied. 

Switching off the scalar self coupling and the Yukawa coupling reduces the fixed points to the usual Banks-Zaks fixed point for the gauge coupling. In the full theory with all three couplings switched on the fixed point generally splits into two distinct fixed points FP$_1^{\pm}$ located in the $(\alpha_g,\alpha_{\lambda})$ plane. In the special case where FP$_1^{+}=$FP$_1^{-}$ there is of course only a single fixed point in the $(\alpha_g,\alpha_{\lambda})$ plane. 
 
There are additional fixed points associated with FP$_2$ above. Looking for zeros of the self coupling beta function at FP$_2$ we find in total two fixed points FP$_2^{\pm}$
\begin{eqnarray}
\alpha_{g,2*} &=& \frac{-b_0}{b_{1}^{\text{eff}}}  \ , \qquad \alpha_{H,2*} =  \frac{c_1 b_0 }{c_2 b_1^{\text{eff}}}  \\ 
\alpha_{\lambda,2*}^{\pm} &=& \frac{b_0 \left( c_2d_2 - c_1 d_3 \pm \sqrt{c_2^2 (d_2^2 - 4d_1d_4)+ c_1^2(d_3^2-4d_1d_5)-2c_1c_2d_2d_3}  \right) }{2b_1^{\text{eff}} c_2 d_1} \ , \qquad \text{FP}_2^{\pm}
\end{eqnarray}
First reality of the fixed points amounts to requiring 
\begin{eqnarray}\label{eq:reality}
l_2 &=& c_2^2 (d_2^2 - 4d_1d_4)+ c_1^2(d_3^2-4d_1d_5)-2c_1c_2d_2d_3 \geq 0
\end{eqnarray}
Then the first fixed point FP$_2^+$ is positive provided 
\begin{eqnarray}\label{eq:FP2p}
b_1^{\text{eff}}>0 \ , \qquad c_2d_2-c_1d_3<0 \  ,  \qquad c_2^2 d_4 + c_1^2 d_5 >0
\end{eqnarray}
while the second fixed point FP$_2^-$ is positive if either
\begin{eqnarray}\label{eq:FP2m}
b_1^{\text{eff}}>0 \ , \qquad c_2d_2-c_1d_3 \leq 0 \qquad \text{or} \qquad b_1^{\text{eff}}>0 \ , \qquad c_2^2 d_4 + c_1^2 d_5 <0 
\end{eqnarray}
Hence if FP$_2^+$ exists then also FP$_2^-$ exists. The opposite might not be the case. Lastly there is of course the special case where $l_2=0$ for which the two fixed points coincide $\text{FP}_2^+=\text{FP}_2^-$ provided $c_2d_2 - c_1d_3 <0$. This concludes our analysis of fixed points for a general gauge-Yukawa theory. In the next section we will investigate the flow of the couplings and whether the fixed points are stable or unstable.

\section{Concluding with the Conformal Phase diagram}
\label{quattro}

Briefly summarizing we found above that there can exist zero, one, two, three or four fixed points for a complete asymptotically free gauge-Yukawa theory with a gauge, Yukawa and scalar self coupling depending on values of the beta function coefficients. The conditions that the beta function coefficients must satisfy are summarized in Table \ref{fig:conditions}. 

\begin{table}
\begin{tabular}{c|c}
  Fixed point & Conditions  \\ \hline\hline
  FP$_1^{\pm}$ & $b_1>0   \ , \qquad  l_1 \geq 0$  \\ \hline 
  FP$_2^+$ & $ b_1^{\text{eff}}>0 \ , \qquad l_2 \geq 0\ , \qquad  c_2d_2-c_1d_3<0 \ ,  \qquad c_2^2 d_4 + c_1^2 d_5 >0$    \\ \hline
  FP$_2^-$  & $b_1^{\text{eff}}>0 \ , \qquad l_2 \geq 0 \ , \qquad  c_2d_2-c_1d_3 \leq 0$  \\
    & or \\ 
    & $b_1^{\text{eff}}>0 \ , \qquad l_2 \geq 0 \ , \qquad  c_2^2 d_4 + c_1^2 d_5 <0 $ \\ \hline
\end{tabular}
\caption{Conditions for the existence of fixed points where $b_1^{\text{eff}} = b_1 - \frac{c_1}{c_2}b_H$, $l_1=d_2^2 - 4d_1d_4$ and $l_2 = c_2^2 (d_2^2 - 4d_1d_4)+ c_1^2(d_3^2-4d_1d_5)-2c_1c_2d_2d_3$. }
\label{fig:conditions}
\end{table}

Having established the existence of all these distinct fixed points we need to discuss along which directions they are attractive or repulsive. First we will start by considering only the gauge and Yukawa couplings and then later study the inclusion of a self coupling. In order to do this we linearise the beta functions around the fixed points and study the eigenvalues and eigenvectors of the matrix 
\begin{eqnarray}
M &=& \left(
\begin{array}{cc}
\frac{\partial \beta_g}{\partial \alpha_g} & \frac{\partial \beta_g}{\partial \alpha_H} \\
\frac{\partial \beta_H}{\partial \alpha_g} & \frac{\partial \beta_H}{\partial \alpha_H}
\end{array}
\right)_{|\alpha_g=\alpha_{g*}, \alpha_H = \alpha_{H*}} 
\end{eqnarray}
The signs of the eigenvalues of $M$ will then indicate whether the associated fixed point is attractive or repulsive along a given eigendirection. Diagonalising $M$ at the fixed point FP$_1$ for which $b_1>0$  we find that the eigenvalues and eigenvectors are 
\begin{eqnarray}
\text{Eigenvalues}\ (M_{\text{FP}_1}) &=& \left(\frac{b_0^2}{b_1} ,\ - \frac{c_1b_0}{b_1} \right)  \\
v_1 &=& \left( 1,0 \right)^T \\
\tilde{v}_1 &=& \left( - \frac{c_2 (b_1-b_1^{\text{eff}})b_0}{b_1 c_1 (b_0 + c_1)} ,1 \right)^T
\end{eqnarray}
The first eigenvalue is always positive. Hence FP$_1$ is attractive in the $ v_1$ direction, i.e. in the $\alpha_g$ direction. It is the fixed point to which the theory flows in the infrared if we also switch off the Yukawa coupling. In this sense it is just the ordinary Banks-Zaks fixed point. The second eigenvalue is always negative. Therefore FP$_1$ is repulsive along the direction $ \tilde{v}_1$ in the ($\alpha_g,\alpha_H$) plane.

Turning to the second fixed point FP$_2$, for which $b_1^{\text{eff}}>0$, we evaluate the matrix $M$ at this fixed point and study its eigenvalues and eigenvectors. They are
\begin{eqnarray}
\text{Eigenvalues}\ (M_{\text{FP}_2}) &=& \left( \frac{b_0^2}{b_1^{\text{eff}}}, \frac{b_1^{\text{eff}} c_1 b_0 +  (b_1 - b_1^{\text{eff}})b_0^2 }{ {b_1^{\text{eff}}}^2 } \right) \\
v_2 &=& \left( \frac{ (-b_1^{\text{eff}}c_1^2 + b_1^{\text{eff}}c_1b_0 - (b_1-b_1^{\text{eff}})b_0^2)c_2 }{ b_1^{\text{eff}} c_1^3 } , 1\right)^T \\
\tilde{v}_2 &=& \left( \frac{(b_1 - b_1^{\text{eff}})(b_0+c_1)c_2b_0}{b_1^{\text{eff}} c_1^3} ,1  \right)^T 
\end{eqnarray}
The first eigenvalue is positive and therefore the fixed point is attractive along $v_2$. In general the second eigenvalue can be either positive or negative. The first term proportional to $b_0$ is always positive but the second term proportional to $b_0^2$ can be of any sign. However we will generally assume that we are investigating a theory for which $b_0$ is a small number number such that the theory sits just below criticality where asymptotic freedom of the gauge coupling is lost. This then ensures that our theory is perturbative and that our analysis in perturbation theory is reliable. We can therefore neglect the order $b_0^2$ squared term and the second eigenvalue will also be positive such that the fixed point is attractive along the eigendirection $\tilde{v}_2$.

\begin{figure}
\begin{minipage}{0.30\textwidth}
\centering
\includegraphics[width=0.9\textwidth]{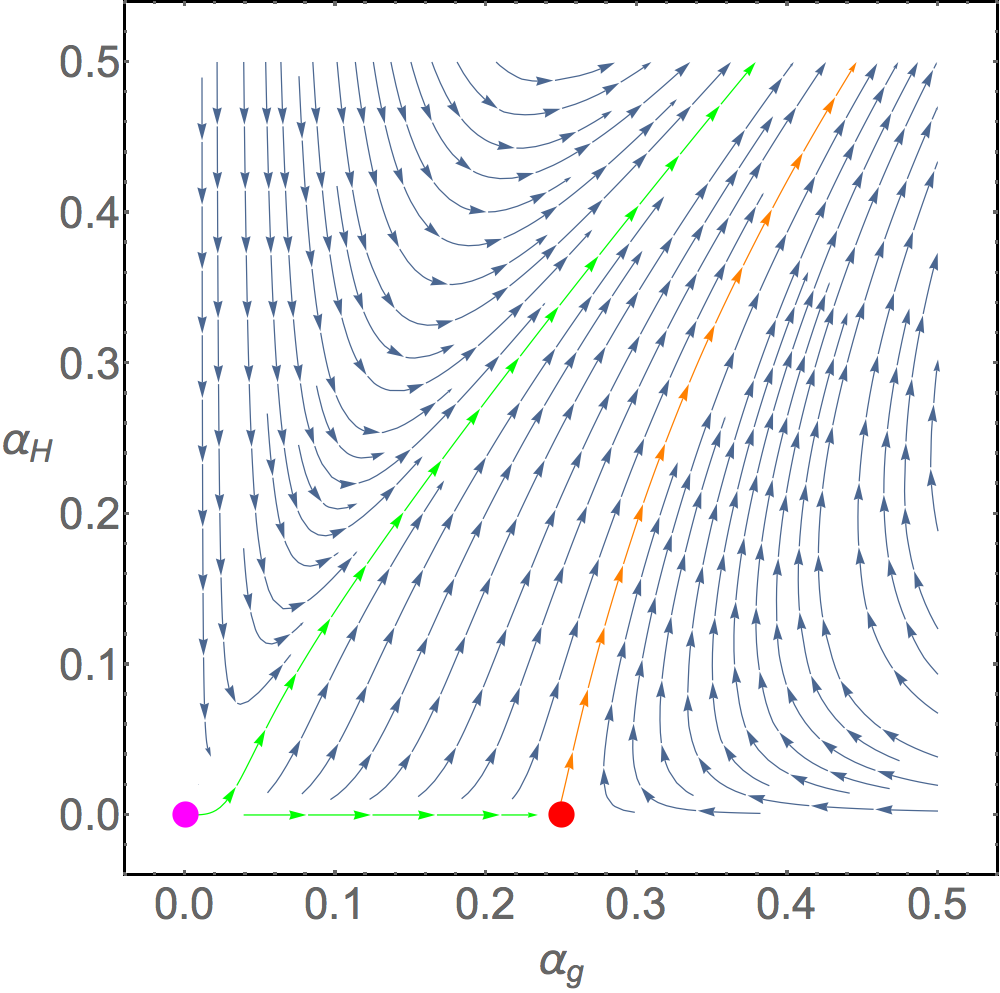}
\end{minipage}
\begin{minipage}{0.30\textwidth}
\centering
\includegraphics[width=0.9\textwidth]{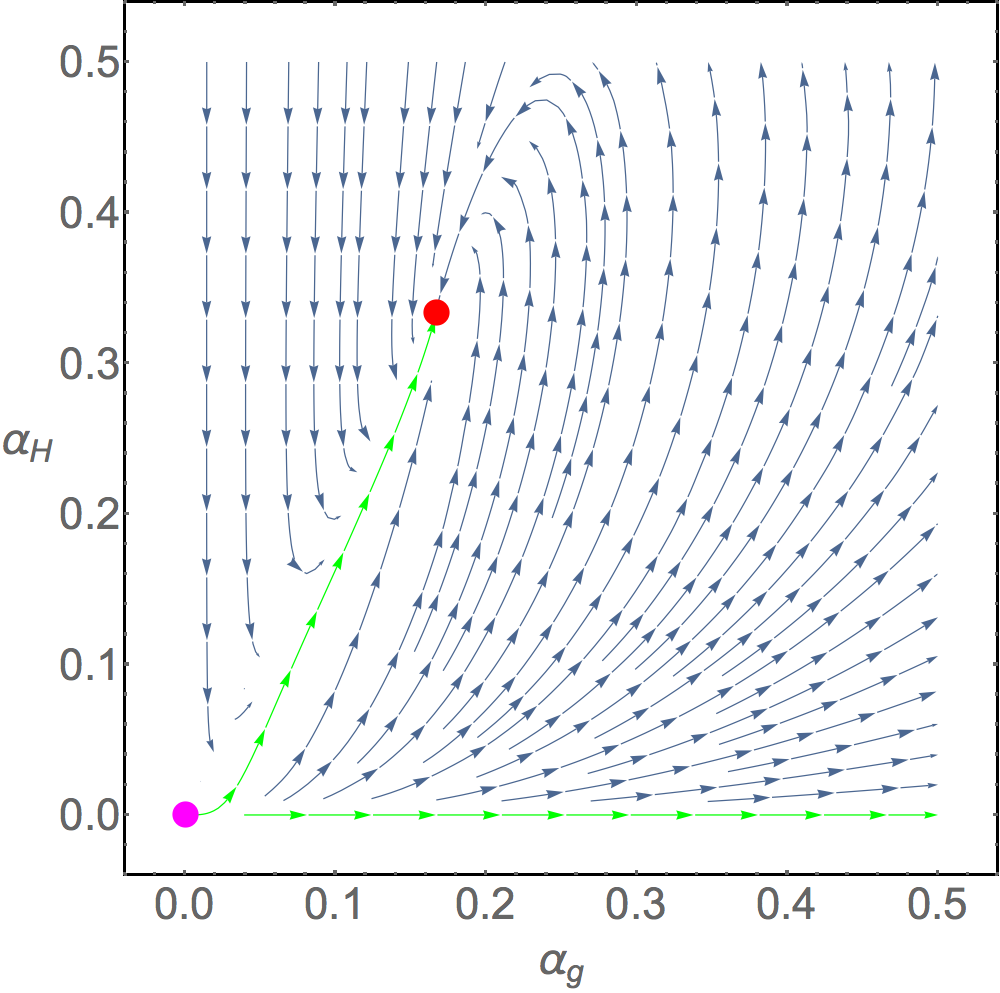}
\end{minipage}
\begin{minipage}{0.30\textwidth}
\centering
\includegraphics[width=0.9\textwidth]{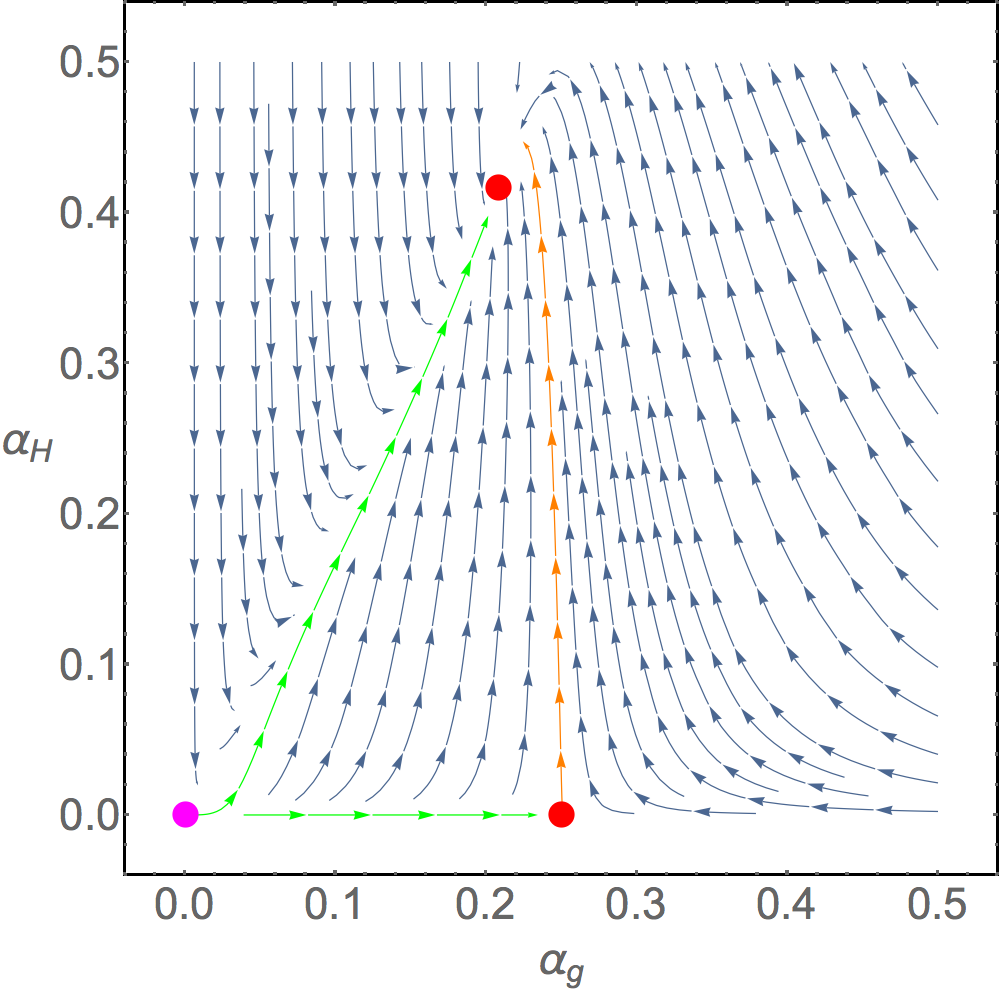}
\end{minipage}
\caption{The flow of the gauge and Yukawa couplings in the case where there is no scalar self coupling. The left panel shows the coupling flow of a theory in which only FP$_1$ exists, the middle panel shows the coupling flow of a theory in which only FP$_2$ exists and the right panel shows the coupling flow of a theory in which both FP$_1$ and FP$_2$ exist simultaneously.}
 \label{fig:gY2loop}
\end{figure}

We choose to plot the flow of the couplings in Fig. \ref{fig:gY2loop} for three different sets of beta function coefficients. One for which only FP$_1$ exists, one for which only FP$_2$ exists and one where both FP$_1$ and FP$_2$ exist simultaneously. The choice of coefficients as well as the eigenvalues and eigendirections of the stability matrix at the fixed points are 

\underline{FP$_1$:}
\begin{eqnarray}
&  b_0=-\frac{1}{2},\ b_1=2,\ b_H=-2,\ c_1 = -2,\ c_2=1, \ b_1^{\text{eff}} = -2 &  \\
& \text{Eigenvalues}\ (M_{\text{FP}_1}) = \left(\frac{1}{8} ,\ - \frac{1}{2} \right)  \ , \qquad  v_1 = \left( 1,0 \right)^T \ , \qquad \tilde{v}_1 = \left(  \frac{1}{5} ,1 \right)^T  & 
\end{eqnarray}

\underline{FP$_2$:}
\begin{eqnarray}
&  b_0=-\frac{1}{2},\ b_1=-1,\ b_H=2,\ c_1 = -2,\ c_2=1, \ b_1^{\text{eff}} = 3 &  \\
& \text{Eigenvalues}\ (M_{\text{FP}_2}) = \left(\frac{1}{12} ,\  \frac{2}{9} \right)  \ , \qquad  v_1 = \left( \frac{1}{3},1 \right)^T \ , \qquad \tilde{v}_1 = \left(  \frac{5}{24} ,1 \right)^T  & 
\end{eqnarray}

\underline{FP$_1$ \& FP$_2$:}
\begin{eqnarray}
& b_0=-\frac{1}{2},\ b_1=2,\ b_H=\frac{1}{5},\ c_1 = -2,\ c_2=1, \  b_1^{\text{eff}} = \frac{12}{5} & \\
&  \text{Eigenvalues}\ (M_{\text{FP}_1}) = \left(\frac{1}{8} ,\ - \frac{1}{2} \right)  \ , \qquad  v_1 = \left( 1,0 \right)^T \ , \qquad \tilde{v}_1 = \left( - \frac{1}{50} ,1 \right)^T & \\
& \text{Eigenvalues}\ (M_{\text{FP}_2}) = \left( \frac{5}{48},\ \frac{115}{288} \right)  \ , \qquad  v_2 = \left( \frac{71}{192} ,1 \right)^T \ , \qquad \tilde{v}_2 = \left( \frac{5}{192}  ,1 \right)^T & 
\end{eqnarray}

From these three plots we see an intriguing phase diagram emerging. Both fixed points FP$_1$ and FP$_2$ are marked by red points with FP$_1$ located on the gauge coupling axis and FP$_2$ located out in the $(\alpha_g,\alpha_H)$ plane. The trivial fixed point is located at the origin and is colored violet. 

First consider the left plot. Here all the trajectories that lie within the boundary marked by the two green and red trajectories (except the red trajectory itself) are the complete asymptotically free trajectories. If both couplings are switched on they will blow up as the infrared regime is approached. If the Yukawa is switched off the gauge coupling just flows to the (Banks-Zaks) fixed point denoted with a red point on the gauge coupling axis. For the trajectories that approach the Banks-Zaks fixed point somewhat closely the gauge coupling will show characteristics of an almost scale invariant system at intermediate scales (i.e. walking dynamics) before blowing up in the infrared (similar behaviour has been observed in semi-simple fermionic gauge theories \cite{Esbensen:2015cjw}).  We also note the special red trajectory for which the fixed point on the gauge coupling axis now acts as an ultraviolet fixed point. Along this red trajectory the Yukawa coupling is asymptotically free while the gauge coupling is asymptotically safe. This is an example of a \emph{safety free} renormalization group trajectory first observed to exist for semi-simple ferminonic gauge theories in  \cite{Esbensen:2015cjw}. 

In the middle plot the Banks-Zaks fixed point for the gauge coupling no longer exists but instead a nontrivial fixed point due to the interplay between the Yukawa and gauge couplings has been generated. The trajectories that lie within the boundary of the two green trajectories are complete asymptotically free and all flow to the fixed point in the deep infrared. Only if the Yukawa coupling is switched off does the gauge coupling blow to large values in the infrared. 

Finally there is the right plot which shows the flow for a theory for which both fixed points exist simultaneously. Again the trajectories that lie within the boundary marked by the two green and red trajectories (expect the red trajectory itself) are the complete asymptotically free trajectories. Here the couplings flow to the nontrivial fixed point in the infrared. Only if the Yukawa coupling is switched off does the gauge coupling flow to the fixed point FP$_1$ on the gauge coupling axis. The trajectories that lie close to FP$_1$ will exhibit near scale invariant behavior for the gauge coupling at intermediate scales before settling at FP$_2$ in the deep IR. Along the special red trajectory the Yukawa coupling is asymptotically free while the gauge coupling is asymptotically safe. Again this a \emph{safety free} trajectory. As the couplings are evolved toward the infrared they both are drawn to the nontrivial infrared fixed point and become again scale invariant. 

Let us now turn our attention to the more involved situation where also the scalar self coupling is switched on. Now in order to study the stability of the fixed points FP$_1^{\pm}$ and FP$_2^{\pm}$ we should linearize the beta function and study the eigenvalues of the matrix
\begin{eqnarray}
M &=& \left( 
\begin{array}{ccc}
\frac{\partial \beta_g}{\partial \alpha_g} & \frac{\partial \beta_g}{\partial \alpha_H} & \frac{\partial \beta_g}{\partial \alpha_{\lambda}} \\
\frac{\partial \beta_H}{\partial \alpha_g} & \frac{\partial \beta_H}{\partial \alpha_H} & \frac{\partial \beta_H}{\partial \alpha_{\lambda}} \\
\frac{\partial \beta_{\lambda}}{\partial \alpha_g} & \frac{\partial \beta_{\lambda}}{\partial \alpha_H} & \frac{\partial \beta_{\lambda}}{\partial \alpha_{\lambda}}
\end{array}
\right)_{|\alpha_g=\alpha_{g*}, \alpha_H = \alpha_{H*}, \alpha_{\lambda} = \alpha_{\lambda*}} 
\end{eqnarray}
The sign of the three eigenvalues of $M$ will determine whether a given fixed point $(\alpha_{g*},\alpha_{H*},\alpha_{\lambda*})$ is attractive or repulsive along an eigendirection. Since both the gauge and Yukawa beta functions do not depend on the scalar self coupling to this order in perturbation theory the first two eigenvalues of $M$ will be identical to the case where the scalar self coupling is switched off as above. 

The first fixed points FP$_1^{\pm}$ are positive provided the constraints in Table \ref{fig:conditions} are satisfied. At theses two fixed points the eigenvalues of $M$ are
\begin{eqnarray}
\text{Eigenvalues} \left( M_{\text{FP}_1^{\pm}} \right) &=& \left(\frac{b_0^2}{b_1}, - \frac{c_1b_0}{b_1}, \pm \frac{\sqrt{l_1}b_0}{b_1} \right) \\
v_1^{\pm} &=& \left(r_1^{\pm},0,1  \right)^T \\
\tilde{v}_1^{\pm} &=& \left( \tilde{r}_1^{\pm}, \tilde{s}_1^{\pm},1  \right)^T  \\
\hat{v}_1^{\pm} &=& \left(  0,0,1 \right)^T
\end{eqnarray}
where $r_1^{\pm},s_1^{\pm}$ and $t_1^{\pm}$ depend on the beta function coefficients and can be found in Appendix \ref{app:coeff}.  The first and second eigenvalues are positive and negative respectively and hence the fixed points FP$_1^{\pm}$ are attractive and repulsive along the associated eigendirections $ v_1^{\pm}$ and $\tilde{v}_1^{\pm}$ respectively. The third eigenvalue is negative at FP$_1^+$ and hence the fixed point is repulsive along the direction $\hat{v}_1^+$ while it is positive at FP$_1^-$ and hence the fixed point is attractive along the direction $\hat{v}_1^-$ Finally we remind ourselves that if the fixed points FP$_1^{\pm}$ exist they exist simultaneously. 

We now move on to study FP$_2^{\pm}$. Here the eigenvalues and eigenvectors are rather complicated expressions of the beta function coefficients so we here provide them to first non vanishing order in $b_0$
\begin{eqnarray}
\text{Eigenvalues} \left( M_{\text{FP}_2^{\pm}}  \right) &=& \left( \frac{b_0^2}{b_1^{\text{eff}}} ,  \frac{ c_1b_0}{b_1^{\text{eff}}} , \pm \frac{\sqrt{l_2} b_0}{c_2 b_1^{\text{eff}} } \right) \\
v_2^{\pm} &=& (r_2^{\pm}, s_2^{\pm}, 1)  \\
\tilde{v}_2^{\pm} &=& (\tilde{r}_2^{\pm}, \tilde{s}_2^{\pm}, 1  )   \\
\hat{v}_2^{\pm}  &=&   (0,0,1) 
\end{eqnarray}
where the coefficients $r_2^{\pm},\ s_2^{\pm},\ \tilde{r}_2^{\pm}$ and $\tilde{s}_2^{\pm}$ are given in Appendix \ref{app:coeff}. The first two eigenvalues are always positive making FP$_2^{\pm}$ attractive along both $v^{\pm}_2$ and $\tilde{v}_2^{\pm}$. The third eigenvalue is negative at FP$_2^+$ and positive at FP$_2^-$. Therefore FP$_2^+$ is repulsive along $\hat{v}_2^+ $ while FP$_2^-$ is attractive along $\hat{v}_2^-$. Note that this makes FP$_2^-$ attractive in all directions. 

There are five different possibilities for these four fixed points to coexist. They are 1) FP$_1^{\pm}$ exist, 2) FP$_2^{\pm}$ exist, 3) FP$_2^-$ exists, 4) FP$_1^{\pm}$ and FP$_2^-$ exist, 5) FP$_1^{\pm}$ and FP$_2^{\pm}$ exist.

\begin{figure}
\begin{minipage}{0.30\textwidth}
\centering
\includegraphics[width=0.9\textwidth]{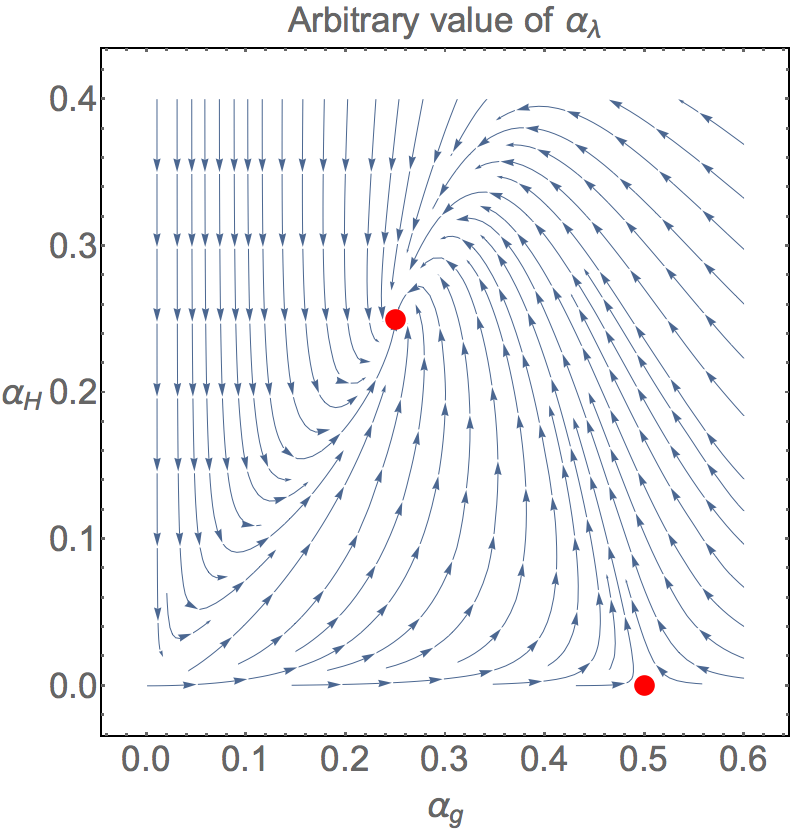}
\end{minipage}
\begin{minipage}{0.30\textwidth}
\centering
\includegraphics[width=0.9\textwidth]{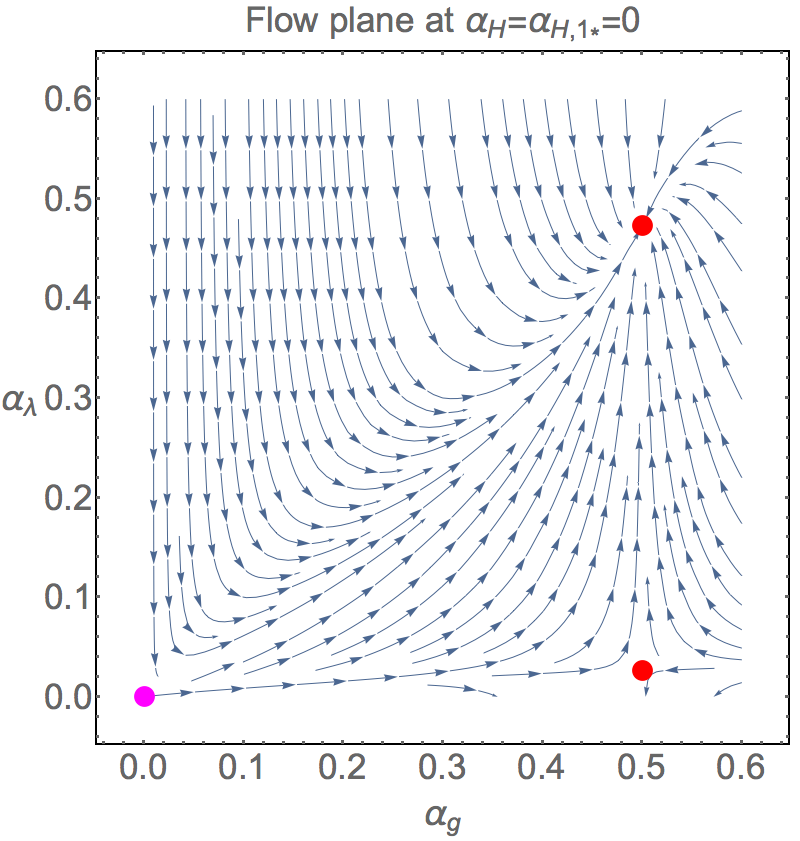}
\end{minipage}
\begin{minipage}{0.30\textwidth}
\centering
\includegraphics[width=0.9\textwidth]{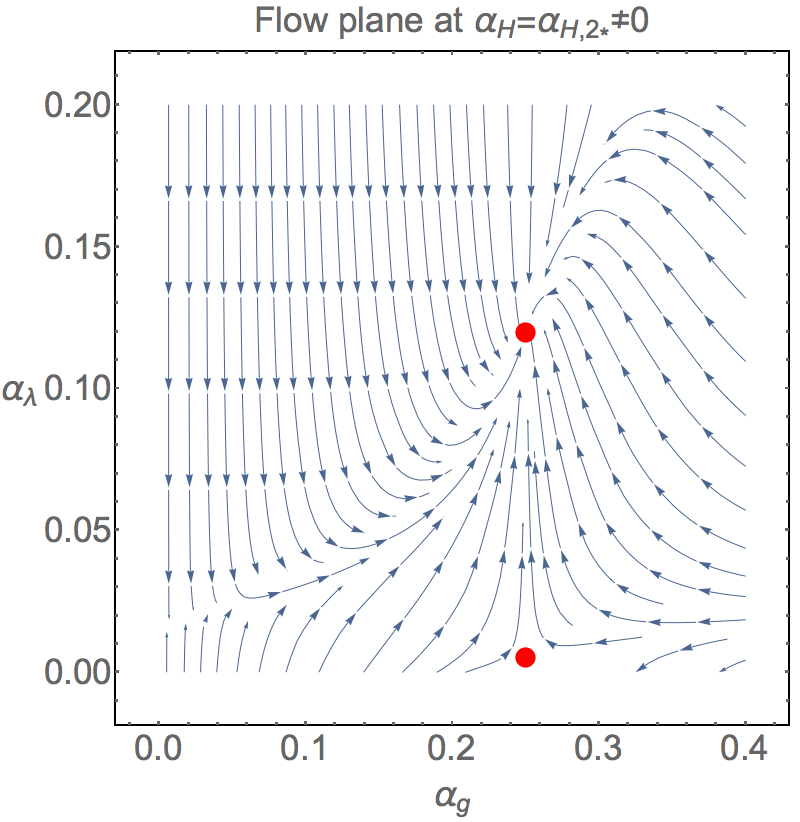} 
\end{minipage}
\\
\begin{minipage}{0.30\textwidth}
\centering
\includegraphics[width=0.9\textwidth]{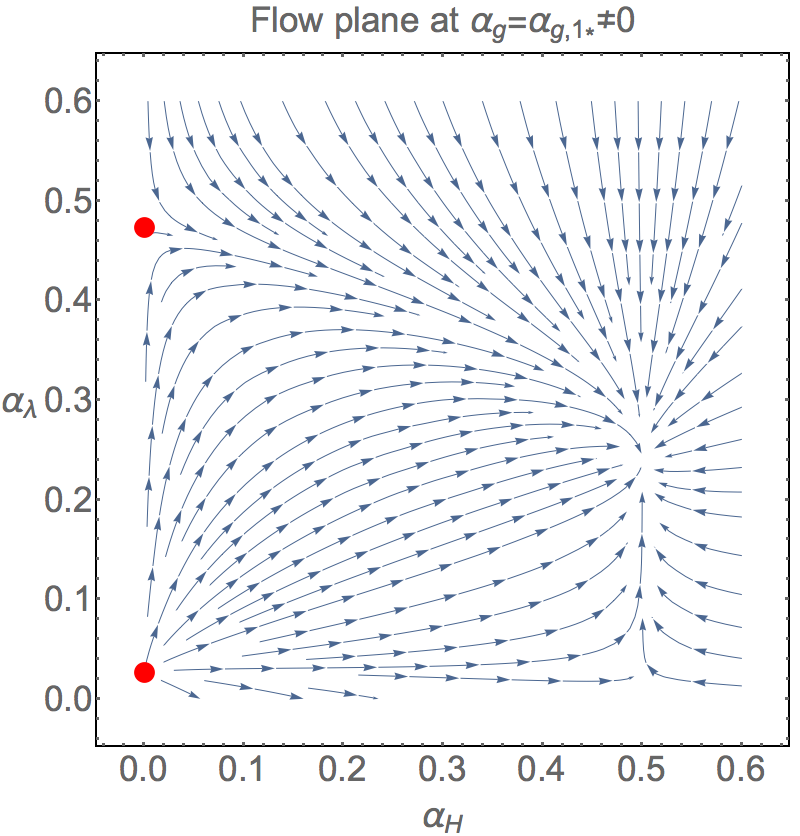}
\end{minipage}
\begin{minipage}{0.30\textwidth}
\centering
\includegraphics[width=0.9\textwidth]{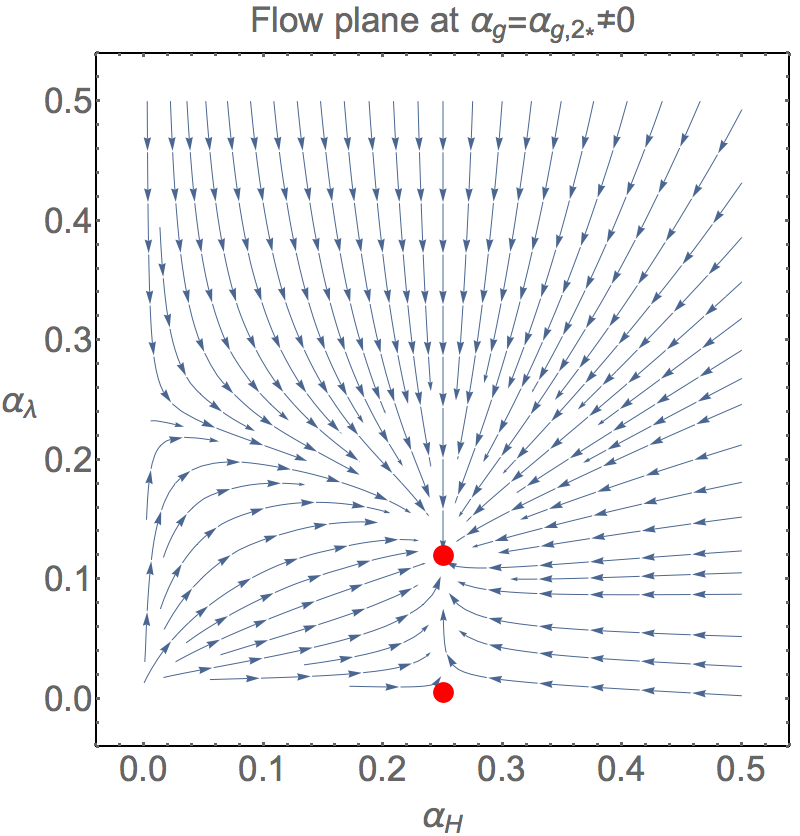}
\end{minipage}
\caption{Flows of the gauge, Yukawa and scalar self couplings projected in the $(\alpha_g,\alpha_H)$ plane (upper left plot), $(\alpha_g,\alpha_{\lambda})$ plane (upper middle and right plots) and $(\alpha_H,\alpha_{\lambda})$ plane (lower left and right plots).}
\label{fig:IRflows}
\end{figure}

We now plot the flow of the couplings for a set of illustrative values of the beta function coefficients in the specific case of 5) where all four fixed points FP$_1^{\pm}$ and FP$_2^{\pm}$ exist simultaneously. The flow can be seen in Fig. \ref{fig:IRflows} for which the values of the beta function coefficients have been chosen to be
\begin{eqnarray}
b_0 &=& -\frac{1}{2} \ , \qquad b_1 =1 \ , \qquad b_H = 1 \ , \qquad c_1 =-1 \ , \qquad c_2 =1 \ , \qquad b_1^{\text{eff}} = 2 \\
d_1 &=& 1 \ , \qquad d_2 =-1 \ , \qquad d_3 = \frac{1}{2} \ , \qquad d_4 = \frac{1}{20} \ , \qquad d_5 = - \frac{1}{25}
\end{eqnarray}

When all three couplings are switched on the flow in coupling space is three dimensional. Since this is difficult to visualize we have plotted only the flow projected into the following planes: i) $(\alpha_g,\alpha_H)$ for the fixed point value of the scalar self coupling $\alpha^*_{\lambda}$ ii) $(\alpha_g,\alpha_{\lambda})$ for the fixed point value of the Yukawa coupling $\alpha^*_{H}$ and iii) $(\alpha_H,\alpha_{\lambda})$ for the fixed point value of the gauge coupling $\alpha^*_{g}$. 

Since the gauge and Yukawa beta functions do not depend on the scalar self coupling to this loop order the coupling flows are identical at the various fixed point values $\alpha_{\lambda,1*}^{\pm}$ and $\alpha_{\lambda,2*}^{\pm}$ of the scalar self coupling. Therefore there is only a single plot in the $(\alpha_g,\alpha_{H})$ plane for an arbitrary value of the self coupling. This is upper left plot in Fig. \ref{fig:IRflows}. The fixed point located on the $\alpha_g$ axis is FP$_1^{\pm}$ while the fixed point located out in the plane is FP$_2^{\pm}$. We have not marked the origin with a violet mark since the depicted flow is for an arbitrary value of the self coupling. It is not necessarily for a vanishing value of the self coupling.

For the flow projected into the $(\alpha_g,\alpha_{\lambda})$ there are two plots located respectively at $\alpha_{H,1*}$ and $\alpha_{H,2*}$. These are the upper middle and upper right plots in Fig. \ref{fig:IRflows} respectively. Since $\alpha_{H,1*} =0 $ we have marked the origin in the upper middle plot with a violet mark which is the ultraviolet trivial fixed point. This is not the case in the upper right plot since this is the flow in the plane at $\alpha_{H,2*} \neq 0$. In the upper middle plot the lower fixed point close to the $\alpha_{g}$ axis is FP$_1^+$ while the fixed point located further out in the plane is FP$_1^-$. In the upper right plot the nontrivial fixed point located close to the $\alpha_g$ axis is FP$_2^+$ while the nontrivial fixed point located further out in the plane is FP$_2^-$. By closer inspection of the upper right plot it seems as if there is an additional fixed point located in the lower left corner on the $\alpha_{\lambda}$ axis that we have missed. However at this point only the gauge and self couplings are at a zero of their beta functions while the Yukawa coupling is not at a zero. In other words there is still a flow out of the plane and hence it is not a real fixed point of the combined gauge, Yukawa and self coupling system.

The flow projected into the $(\alpha_H,\alpha_{\lambda})$ plane at the two different locations $\alpha_{g,1*}$ and $\alpha_{g,2*}$ are plotted in the lower left and right plots respectively. Again we have not marked the origin violet since in both cases the flow is plotted for a nonvanishing value of the gauge coupling. In the lower left plot the lower fixed point is FP$_1^+$ while the upper fixed point is FP$_1^-$. In the lower right plot the fixed point located close to the $\alpha_H$ axis is FP$_2^+$ while the fixed point located further out in the plane is FP$_2^-$. Again it seems as if we have missed two fixed points in the two lower plots. However similar to above these only correspond to zeros of the Yukawa and self coupling beta functions but not simultaneously of the gauge beta function. Hence at these locations there is still a flow in a direction out of the plane. 

There are a number of important conclusions that arise by close inspection of the possible flows. First there is the type of flow which we originally set out to study. This is the flow where all couplings are asymptotically free and then flow to a nontrivial infrared stable fixed point. This is the fixed point FP$_2^-$ which is infrared attractive in all directions (as noted above). 

However there is additional phase structure that we can uncover. For instance in the upper left plot there is a special trajectory that connects the lower fixed point with the fixed point located out in the plane. Along this trajectory the lower fixed point now acts as an ultraviolet fixed point while the other fixed point is an infrared fixed point. For the Yukawa coupling the ultraviolet fixed point is trivial while for the gauge coupling it is nontrivial. Hence the Yukawa coupling is asymptotically free while the gauge coupling is asymptotically safe. The scalar self coupling along this flow is constant and does not run. 

Also along the trajectory connecting the lower red fixed point with the upper red fixed point in the upper middle and upper right plots the lower fixed point acts now as a nontrivial ultraviolet fixed point while the upper fixed point is a nontrivial infrared fixed point. In the upper middle plot the Yukawa coupling vanishes along the entire flow while in the upper right corner it assumes a constant nonvanishing value along the entire flow. The gauge coupling assumes a constant nonvanishing value along both flows. Hence only the scalar self coupling runs between two fixed points while the gauge and Yukawa couplings are constant. Similar types of dynamics can be observed in the lower left and right plots where there are trajectories that connect the two nontrivial fixed points where one acts as an ultraviolet fixed point while the other acts as an infrared fixed point. Along these special trajectories only the scalar self coupling runs.

We therefore conclude that there are trajectories in coupling space in which 1) all three couplings flow nontrivially between two fixed points 2) only two couplings flow nontrivially between two fixed points with the remaining coupling being constant and 3) only a single coupling flows nontrivially between two fixed points with the remaining two being constant.

\vskip .4cm
The analysis performed here elucidates the immense richness of the conformal structure of gauge-Yukawa theories. We focussed here on a  time-honoured class of such theories known as complete asymptotically free.  Here  gauge, Yukawa and scalar couplings achieve an ultraviolet  noninteracting fixed point. Our work, for the first time, investigated the important infrared conformal structure of these theories.  We revealed the occurrence of several novel conformal phenomena associated to the emergence of different types of interacting fixed points. The applications to beyond standard model physics are limitless ranging from the construction of potential new classes of dark matter to inflationary models as well as (composite) dynamics featuring elementary scalars that are fundamental according to Wilson.

\acknowledgments
 The $\mathrm{CP}^3$-Origins centre is partially funded by the Danish National Research Foundation, grant number DNRF90.

\appendix

\section{Eigendirections}\label{app:coeff}

Here we provide the coefficients that enter in the expression for the various eigendirections
\begin{eqnarray}
r_1^{\pm} &=&  \frac{2d_1(b_0\mp \sqrt{l_1})}{l_1\pm d_2\sqrt{l_1}} \\
\tilde{r}_1^{\pm} &=&  b_0c_2  (b_1-b_1^{\text{eff}})  \frac{2d_1 (c_1 \pm \sqrt{l_1})}{(b_1 c_1^2 d_3 +  b_0(b_1c_1 d_3\mp  b_1c_2\sqrt{l_1} \pm b_1^{\text{eff}} c_2 \sqrt{l_1} ) ) (d_2 \pm  \sqrt{l_1}) } \\
\tilde{s}_1^{\pm} &=& - b_1 c_1 (b_0 + c_1) \frac{2 d_1 (c_1 \pm \sqrt{l_1} ) }{(b_1 c_1^2 d_3 + b_0 (b_1c_1d_3 \mp b_1c_2 \sqrt{l_1} \pm b_1^{\text{eff}} c_2 \sqrt{l_1} )) (d_2 \pm \sqrt{l_1})} \\
r_2^{\pm} &=&  - \frac{2c_2d_1}{c_2d_2 - c_1d_3 \pm \sqrt{l_2}} \left[ 1 \mp \frac{c_2^2 d_2 + c_1(d_3^2 - 4 d_1 d_5) - c_2 (c_1 d_3 + d_2 d_3 \mp \sqrt{l_2}) \mp d_3 \sqrt{l_2}}{ (c_2d_2 - c_1 d_3 \pm \sqrt{l_2}) \sqrt{l_2}} b_0 \right]  \\
s_2^{\pm} &=& \frac{2c_1 d_1}{c_2d_2 - c_1 d_3 \pm \sqrt{l_2} }  \left[1 \pm \frac{(c_2d_2^2 + c_1^2 d_3 - 4 c_2 d_1 d_4 - c_1(c_2 d_2 + d_2 d_3 \pm \sqrt{l_2}) \pm d_2 \sqrt{l_2}  )  b_0}{(c_2d_2 - c_1 d_3 \pm \sqrt{l_2}) \sqrt{l_2}} \right]  \\
\tilde{r}_2^{\pm} &=& \frac{2 (b_1 - b_1^{\text{eff}})c_2d_1 (c_1 c_2 \mp \sqrt{l_2})}{ b_1^{\text{eff}} c_1^2 (c_2d_2d_3 - c_1d_3^2 + 4 c_1 d_1 d_5 \pm d_3 \sqrt{l_2} ) }b_0 \\
\tilde{s}_2^{\pm} &=& \frac{2d_1}{c_2d_2d_3 - c_1d_3^2 \pm d_3 \sqrt{l_2}} \nonumber \\
&& \left[ c_1c_2 \mp \sqrt{l_2} \mp \frac{(b_1-b_1^{\text{eff}})c_2 (-c_2d_2^2 - c_1^2 d_3 + 4 c_2 d_1 d_4 + c_1(c_2 d_2 + d_2 d_3 \pm \sqrt{l_2}) \mp d_2 \sqrt{l_2})\sqrt{l_2} }{b_1^{\text{eff}} c_1^2 (c_2d_2d_3 - c_1 d_3^2 + 4c_1d_1d_5 \pm d_3 \sqrt{l_2} ) }b_0  \right] \nonumber\\
\end{eqnarray}



\end{document}